\documentclass[]{spie}
 
\usepackage{textcomp,amsmath,amsfonts,amssymb,gensymb}
\usepackage{graphicx}
\usepackage[colorlinks=true, allcolors=blue]{hyperref}
\usepackage{comment}
\usepackage{multirow}

\newcommand{\micron}{\mbox{$\mu$m}}

\newcommand{\KEN}{\texttt{KEN}}
\newcommand{\KSN}{\texttt{KSN}}
\newcommand{\TCUT}{\texttt{TCUT}}
\newcommand{\act}{\texttt{\_A}}
\newcommand{\darkref}{\texttt{\_R}}
\newcommand{\phan}{\texttt{\_P}}
\title{Tracing the Boundary of the South Atlantic Anomaly Region with SPHEREx Transient Flagging Rates}

\author[a]{Chi H. Nguyen}
\author[a,b]{James Bock}
\author[c]{Sean Bryan}
\author[a]{Walter R. Cook}
\author[b]{Brendan Crill}
\author[b]{C. Darren Dowell}
\author[a,b]{Olivier~Dor\'{e}}
\author[b]{Beth Fabinsky}
\author[d]{Candice Fazar}
\author[a]{Howard Hui}
\author[a]{Phillip Korngut}
\author[a]{Steve Padin}
\author[d]{Michael Zemcov}

\affil[a]{California Institute of Technology, Pasadena CA 91125, USA}
\affil[b]{Jet Propulsion Laboratory, California Institute of Technology, Pasadena CA 91109, USA}
\affil[c]{Arizona State University, Tempe AZ 85287, USA}
\affil[d]{Rochester Institute of Technology, Rochester NY 14623, USA}

\authorinfo{Send correspondence to C.H.Nguyen: chnguyen@caltech.edu}

\begin{document} 
\maketitle

% ABSTRACT
\begin{abstract}
The NASA SPHEREx satellite was launched in 03/2025 to survey the full sky between 0.75 - 5.0 $\micron$. 
The image processing of SPHEREx H2RG detectors includes real-time flagging of transient events during integration. 
SPHEREx follows a polar orbit passing over the South Atlantic Anomaly (SAA) zone multiple times daily, where the transient counts reach as high as $>$80\% of pixels, depleting the pixel inventory for further analysis. 
The science pipeline flags all exposures taken while in the SAA for re-observation, as well as any with $>$10\% transients regardless of the spacecraft location. 
After six months, $\sim$~1.5\% of the exposures outside of the pre-defined SAA zone were flagged. 
While some exposures are adjacent to the edge of the SAA, the majority appear to trace the geomagnetic field near the poles. 
Given the importance of the SAA to designing and operating space instruments, in this paper we present the SAA boundary traced by SPHEREx transient flags as a reference for future missions.
\end{abstract}

% KEYWORDS

\section{INTRODUCTION} \label{S:intro}
The South Atlantic Anomaly (SAA) is a region in the Earth's magnetic field where the observed field strength drops by at least a factor of 2-3$\times$ \cite{PavonCarrasco2016}.
As a result of the weakened field, the van Allen radiation belt dips to much lower altitude in the SAA, which allows highly energetic particles to penetrate deeper into the atmosphere directly and affects the communication and navigation of satellites passing through the region \cite{Deme1999, Finlay2020}.
The exact location of the SAA varies; historically, it has been observed to shift from South Africa to South America at the rate of $\sim$tenth of a degree per year \cite{Stassinopoulos2015}.
The SAA is of great interest to research in Earth science, heliophysics, space weather, and satellite operation.
Monitoring the SAA and its temporal changes is crucial to understand the origin and the evolution of the Earth's magnetic field.
For example, modeling efforts to explain the SAA's existence point to a pending magnetic field reversal \cite{Finlay2016, Laj2015, DeSantis2013}.
At the same time, satellite operation in low Earth orbit (LEO) depends heavily on knowledge of the SAA boundary and its effects on electronics systems \cite{Bezerra2009} especially during geomagnetic storms, when the coupling between the magnetosphere-ionosphere can be significantly disturbed.
Given its role, the SAA has been the subject of multiple studies using ground-based and space-borne instruments, either by directly measuring the strength of the magnetic field lines or by indirectly mapping the intensity of particle fluxes \cite{Bartocci2025, Ginisty2024, Ginisty2023} and counting anomaly events on satellites \cite{Finlay2020}.
Archaeomagnetic investigations into the historic boundaries of the SAA include modeling of naval ship navigation logs \cite{Schanner2023} and radioisotope dating in pottery \cite{Gomez-Paccard2025}.

SPHEREx (Spectro-Photometer for the History of the Universe, Epoch of Reionization, and ices Explorer) is a NASA Medium Explorer~(MIDEX) satellite currently operating in low Earth orbit (LEO) that will perform four all-sky near-infrared (NIR) spectral surveys between 0.75 -- 5.0 \micron\cite{Bock2025}.
The instrument and the survey plan is constructed to address a broad range of astrophysical questions through low-resolution spectroscopy over the nominal two-year mission. 
Launched in March 2025, SPHEREx completed its first year of data collection and produced two full-sky surveys by mid-May 2026.
The science instrument comprises six HAWAII-2RG (H2RG) detectors that continuously collect spectral images of the sky in 115-second integrations.

Due to the large volume of the survey data, the raw detector data are processed on-board and only the photocurrent images at the end of the integration are downlinked \cite{Nguyen2025}.
A real-time flagging algorithm is implemented to identify pixels that have been affected by transient events during the observation, as well as detector- and electronic-related effects like saturation \cite{Zemcov2016}.
The percentage of pixels flagged as transient events in an observation taken while SPHEREx is in the SAA (as determined by the ground track latitude/longitude) is exceedingly high, and the remaining good pixels are generally not sufficient for astrophysics analysis. Therefore, these observations are preemptively excluded from the high level data processing and the sky targets of these observations are re-observed.
During in-orbit commissioning (IOC), the transient percentage in the SAA observations were found to fall in the range of $10 - 80\%$ of the total pixel count in every detector.
Most notably, the flagging fractions are not randomly distributed, but correlate with the locations of the spacecraft within the SAA.
Furthermore, $1 - 2\%$ of observations outside of the SAA (refered to hereafter as ``non-SAA'' observations) are found to have transient fraction $\gtrsim10\%$, high enough to render them challenging to process.
These observations cluster at the edge of the SAA and near the terrestrial poles, aligned with the geomagnetic North and South poles.
Following IOC, a new step was added in the pipeline to flag the non-SAA exposures with high transient counts for re-observation.

Although SPHEREx is designed to be an astrophysics mission, the survey data provide continuous global measurements of the NIR airglow in the upper atmosphere \cite{Hui2026b} which are of great interest to atmospheric science and heliophysics.
In this paper, we present another case in which SPHEREx data can be useful beyond the field of astrophysics.
While the SAA-flagged data do not meet the standard for astrophysical studies, the transient rates recorded in these observations present a coherent map of the SAA as well as the geomagnetic poles from LEO.
Coincidentally, SPHEREx was launched during the declining phase of the Solar maximum, and subsequently experienced various geomagnetic disturbance levels just in its first year in orbit, enabling assessment of the SAA in different geomagnetic conditions.
In this work, we report the state of SPHEREx transient flagging and the SAA structure as traced by the transient rates, using 206,379 observations collected between the start of May 2025 (week 18 of 2025) and the end of April 2026 (week 18 of 2026).

The paper is structured as follows.
In section \ref{S:transient_saa}, we review SPHEREx real-time transient flagging algorithm and how the SAA is flagged.
The one-year data of transient flagging is presented in section \ref{S:rate_in_survey}.
We map the transient flagging rates to the SAA in section \ref{S:discussion}, and discuss notable trends in different geomagnetic conditions.
A summary is included in \ref{S:summary}.

\section{TRANSIENT AND SAA FLAGGING IN SPHEREx}\label{S:transient_saa}

\subsection{SPHEREx Highlights}
SPHEREx is designed to map the sky in 102 wavebands between $\lambda = 0.75 - 5.0$ \micron~ to study the origin and physics of cosmic inflation, refine constraints on galaxy formation, and search for ices in the Milky Way \cite{Bock2025}.
In addition, SPHEREx will also produce a legacy catalog of millions of galaxy redshifts and low-resolution spectra of nearby stars.
The calibrated data are processed by the SPHEREx Science Data Center at Caltech/IPAC and available from the Infrared Science Archive\footnote{IRSA web portal for SPHEREx data: \protect\url{https://irsa.ipac.caltech.edu/Missions/spherex.html}\cite{Akeson2025}} (IRSA).
SPHEREx was launched in March 2025, with in-orbit commissioning (IOC) following immediately between March and April 2025.
The science data collection started in May 2025 and two full sky surveys were completed by mid-May 2026.

SPHEREx orbits along the day-night terminator line at a mean altitude of 650 km\footnote{SPHEREx altitude ranges from $\sim$645 - 680 km} in a sun-synchronous polar orbit with period of $\sim98$ minutes, passing over the terrestrial poles as well as the SAA multiple times a day. 
Except for small time gaps between exposures to change the telescope pointing and for scheduled data downlink, SPHEREx continuously scans the sky as it orbits the Earth.
By orbiting along the terminator line, SPHEREx science instrument can collect sky images without interruption on the night side while the solar panel is constantly illuminated by the Sun to maintain stable power.
An advantage of the polar orbit is that the same sky regions at the North and South ecliptic poles (NEP/SEP) are observable by SPHEREx multiple times daily, allowing these two regions to have significantly more observations that can be averaged for better sensitivity.
The observations of the NEP and SEP are considered ``deep field'' data, and the rest of survey is referred to as ``all sky'' hereafter.

SPHEREx consists of a wide field-of-view telescope coupled to six H2RG detectors split into two focal plane assemblies (FPA) covering 0.75 - 2.5 $\micron$ (detectors 1 to 3), and 2.5 - 5 $\micron$ (detectors 4 to 6) \cite{Korngut2026}.
Each detector comprises a 2048$\times$2048 pixel array with 2040$\times$2040 optically active pixels, and 4 rows/columns of optically dark reference pixels on every edge of the detector.
A linear variable filter is mounted in front of each detector to vary the spectral transmission along one dimension of the detectors\cite{Hui2026}.
The six H2RG are simultaneously integrated for 115 seconds before reset, sufficient to reach the required source sensitivity for the galaxy redshift and ices surveys \cite{Bock2025}.
To minimize instrument noise, in particular the frequency-dependent ($1/f$) noise, the detectors are read multiple times during the 115-second integration as the charges accumulate (a technique known as sampling-up-the-ramp or SUR), with row chopping and multiple samplings of optically dark reference pixels interwoven in between readings of optically active pixels.
Additionally, random amplifier drift is measured four times at the beginning of every pixel row and saved into the data frame as ``phantom pixels'' \cite{Nguyen2025, Heaton2023}.

\subsection{Real-time Transient Detection}
Given the large volume of data, individual SUR readings are not saved during integration.
Instead, the photocurrent slope (sky flux) in every pixel is computed by the on-board instrument control electronics (ICE) software in real-time during an integration, and only the final SUR slope image is downlinked for science \cite{Zemcov2016}.
A real-time flagging mechanism at the pixel level is required to detect transient events, as well as various detector effects like pixels approaching nonlinearity or saturation \cite{Zemcov2016}.
Additional electrical effects like persistence and charge blooming are identified during ground processing of the data, taking into account the existing real-time SUR flags.
Transient events include cosmic rays, satellite trails, as well as optical effects like flares from near-field satellites.
The SUR flags are saved for every pixel in an array-like format similar to the sky image, and are included in the calibrated data (together with other flags from later calibration steps) \cite{Akeson2025}.

A version of the transient detection algorithm was previously published with the real-time photocurrent estimate framework \cite{Zemcov2016}.
Presented below is the updated algorithm as used in flight data.
As the integration proceeds, more electrons accumulate in every pixel and the pixel is read out approximately every 1.5 seconds. 
After each read, the transient detection algorithm compares the current pixel value with a prediction based on the prior four reads.
If the difference between prediction and actual value is statistically unlikely for a given pixel, the transient flag is set and only the photocurrent calculated from reads \textit{before} the transient event is saved for this pixel.
For every integration, the first two reads after resetting the detector are not used in the SUR and flagging algorithm to avoid post-reset settling effect.
The third sample is the first read to be used in SUR.
Given that at least four preceding consecutive reads are needed to apply the transient condition (Equation \ref{Eq:transient_condition}), the transient detection is only in effect starting at the 7th sampling following the detector reset.
The condition for a transient detection is:
\begin{equation}
    (2y_i - 2y_{i-1} - y_{i-2} + y_{i-4})^2 > 4\times \KEN + \KSN\times (y_{i-1}-y_{i-4} - 3m)
    \label{Eq:transient_condition}
\end{equation}
where index $i$ refers to the current readout frame; $[y_{i-1} + (y_{i-2} - y_{i-4})/2]$ is an optimized predictor for the current read value $y_i$; \KEN~(transient white noise) and \KSN~(transient shot noise) are configurable parameters that can be set individually for each detector and for each of the three pixel types (optically active pixels, optically dark reference pixels, and phantom pixels).

The first term on the right-hand side of Equation \ref{Eq:transient_condition} is designed to look for statistical outliers for an approximately gaussian distribution of noise (applicable at low photocurrent), and the second term loosens the threshold to account for Poisson noise at higher photocurrent. 
$m$ is the amplifier drift rate measured during the prior exposure.
The unit of $y$ is ADU; for SPHEREx, 1 ADU corresponds roughly to $\sim 0.8 - 1.0$ electron depending on the detectors.
The in-flight values of \KEN~and \KSN~are determined from pre-flight laboratory characterization and fixed for the entire survey.

The SUR photocurrent noise decreases with longer integration time until the detector reaches the $1/f$ read noise floor \cite{Nguyen2025}. Therefore, the slope of a pixel flagged near the beginning of an observation can have much higher noise than the rest of the array due to its shorter effective integration time.
However, the on-board flight software does not keep track of when a transient is detected for every event.  
Another configurable parameter, \TCUT, specifies a frame index that sub-divide the transient events into ``early'' and ``late'' groups.
If a transient event is detected in a pixel before frame index \TCUT, an additional ``early'' flag is set for this pixel to identify that its noise is potentially higher than those in the late group.

In Table \ref{tab:flight_params}, we list the values of the transient detection parameters in the sky survey.
The information is also included in the SPHEREx Explanatory Supplement \cite{Akeson2025}.
For the current value of \KEN\act, the transient detection threshold is $\geq \frac{1}{2}\sqrt{4\times \KEN\act} = 141\ \textrm{ADU} \approx 100$ e-.  
An early transient is flagged if detected before $\TCUT = 38$, or approximately halfway through the 115-s integration.

\begin{table}
\begin{center}
\renewcommand{\arraystretch}{1.25}
\begin{tabular}{ccll}
\hline
\hline
Parameters  & Value  & Unit     & Note\\
\hline
\KEN\act      & 20000  & ADU$^2$  & White noise variance threshold, optically active pixels\\
\KEN\darkref      & 10000  & ADU$^2$  & White noise variance threshold, reference pixels\\ 
\KEN\phan      & 500    & ADU$^2$  & White noise variance threshold, phantom pixels\\
\hline
\KSN\act      & 100    & ADU  & Shot noise threshold, optically active pixels\\
\KSN\darkref      & 0      & ADU  & Shot noise threshold, reference pixels\\ 
\KSN\phan      & 0      & ADU  & Shot noise threshold, phantom pixels\\
\hline
\TCUT        & 38     & -    & SUR sample index threshold for early transient\\
\hline
\end{tabular}
\end{center}
\caption{Transient detection parameters.\label{tab:flight_params}}
\end{table}

\subsection{SAA Flagging in the Survey Plan}\label{S:SAA}
To simplify the operation of SPHEREx, the instrument does not stop collecting data near or in the SAA.
Instead, we apply a pre-determined flag in the survey plan to identify exposures taken when the spacecraft is in the SAA and remove them from the data processing on the ground.

The specific target pointing at a given time is optimized by a survey planning software (SPS) \cite{Bryan2025} taking into account the predicted location of the spacecraft, the state of the coverage-to-date, as well as the thermal and stray light constraints from the Sun, the Moon, and the Earth.
The targeting points are selected on a 3.5-day cadence, hereafter referred to as a ``planning period''\footnote{The bi-weekly planning period can be determined by a prefix in the observation identification (OBSID) found in the header of the data file: \texttt{$<$year$>$W\#\#\_\$$<$version letter$>$}, where \texttt{\#\#} is the two-digit zero-padded index of the week in the year starting from 1 and \texttt{\$} = 1 or 2, the first or second half of the week respectively. Example: \texttt{2025W18\_1B}.}.
Using the existing location data of the spacecraft as tracked by an onboard GPS\footnote{The ground-tracked coordinates can be found in the header of the data file as \texttt{SGT\_LAT\_MIDPT} and \texttt{SGT\_LON\_MIDPT}.}, the future locations of the spacecraft in a planning period can be predicted accurately in advance.
Observations forecasted to be taken when the spacecraft is in the boundary of the SAA at SPHEREx orbital altitude (Figure \ref{fig:heatmap_all}) are identified in advance.
The observations labeled as ``SAA'' are not passed through the calibration steps, based on the expectation that high number of transient detections in the SAA result in too few good pixels available for calibration.

\subsection{Re-observation of High Transient Data}
Early in the survey, we found that the transient rates persist up to $10\%$ for a non-negligible fraction of observations taken outside of the SAA.
The majority of these were taken when the spacecraft was approaching and leaving the SAA, but many were not in the vicinity of the SAA.
These exposures subsequently fail the calibration steps and quality check, prompting an additional cut to be added to the data processing.
Non-SAA observations are removed from further processing if they are \textit{not} deep field targets and their transient hit counts are above 435,000 in any one detector ($\sim10\%$ of pixel count per detector) \cite{Akeson2025}.
The NEP/SEP coverage is dense and there is sufficient overlapping in ecliptic coordinates between the pointing targets in the deep fields that there is no need to successfully observe every target position.
The pixel threshold is selected based on the performance of the data pipeline when applied on early survey data.

Following the SAA and high transient cuts, the good deep and all sky exposures undergo a series of calibration and quality checks before calibrated images in sky flux unit are produced \cite{Akeson2025}.
Despite not being used by the science pipeline, the high transient and SAA data as well as their corresponding SUR flags are archived internally to the mission team to monitor the health and performance of the instrument.
To ensure full sky coverage, the pointing targets of the removed exposures are sent back into the observing queue to be re-observed.

\section{TRANSIENT RATES IN ONE YEAR SPHEREx DATA}\label{S:rate_in_survey}

\begin{figure}[htb]
	\centering
    \vspace{5pt}
	\includegraphics[width=0.75\textwidth]{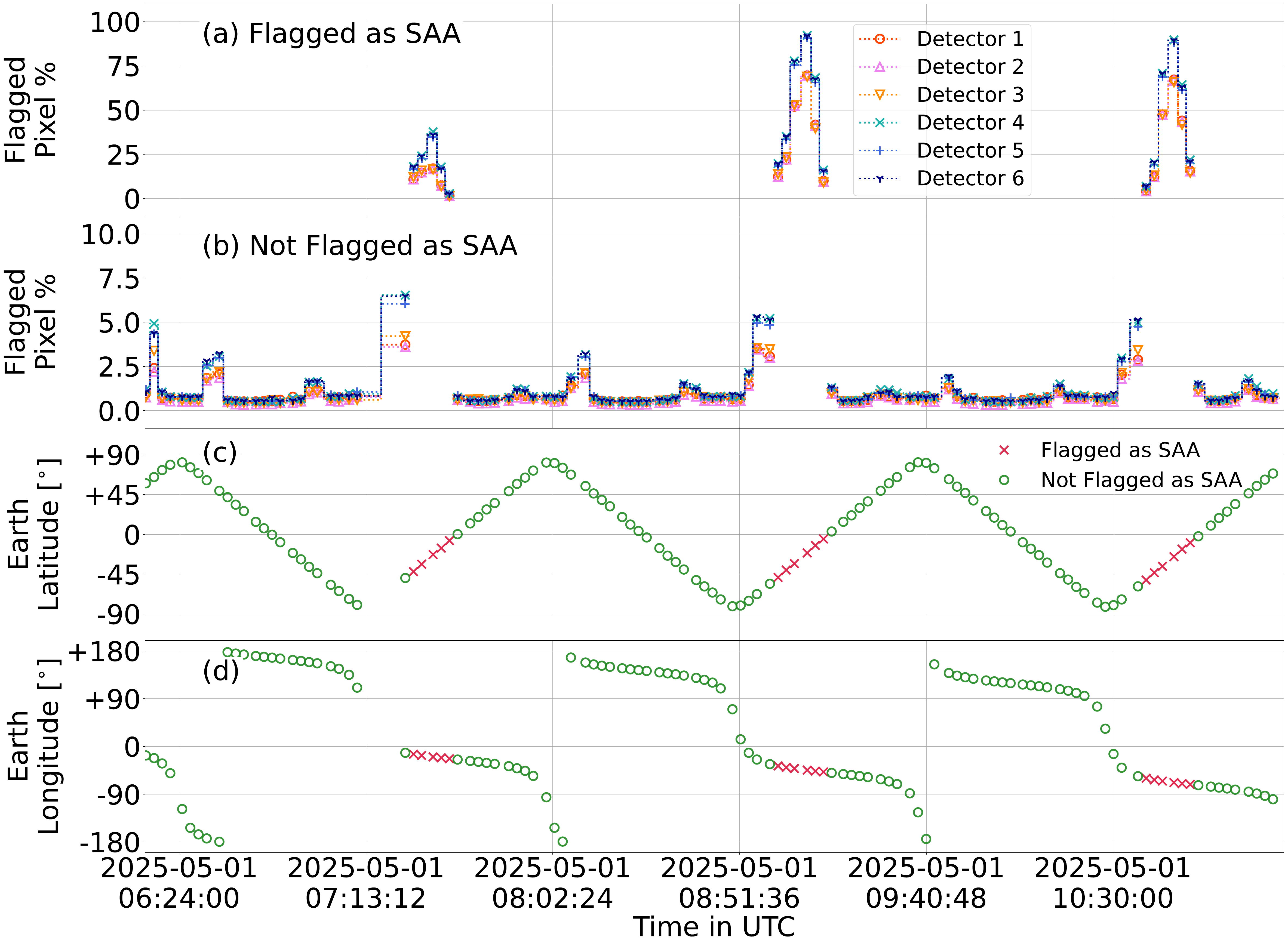}
    \vspace{5pt}
	\caption{Time series of a six-hour period to illustrate the correlation between the transient rates and the ground-tracked locations of SPHEREx.
    (a) Transient rates in observations taken in SAA, (b) transient rates in observations not flagged as SAA, (c) SPHEREx latitude, and (d) SPHEREx longitude at the time of each observation.
    The rates are reported as percentage of flagged pixels relative to the total number of pixels in an array, and are separated to emphasize the range of SAA observations, reaching as high as $>80\%$ transients, in comparison to the non-SAA observations rate of $\leq 10\%$.
    The SAA observations are pre-determined using prediction of the spacecraft locations on a 3.5-day cadence, denoted in the latitude and longitude plots by the red x marks.
    A non-negligible number of non-SAA observations exhibit elevated transient rates, either due to approaching the SAA or by passing over the geomagnetic poles.
    Additionally, detectors 1 - 3 (2.5$\micron$ cutoff H2RG) consistently record fewer transient events than detectors 4 - 6 (5$\micron$ cutoff H2RG).
    \label{fig:timeseries_zoomin}}
\end{figure}

In Figure \ref{fig:timeseries_zoomin}, we plot the time series of the transient rates as the percentage of the total pixel counts for every detector, as well as the ground-tracked latitudes and longitudes of the spacecraft in a six-hour period. 
The observations marked as SAA based on the predicted spacecraft location are highlighted in the longitude and latitude plots, and their transient rates are plotted separately from non-SAA observations to emphasize the dynamic range of the transient rates in the SAA.

Two prominent trends can be inferred from the time series.
First, while the location-based flagging of the SAA accurately captures the observations with exceedingly high transient counts, many exposures in the non-SAA group exhibit an elevated level of transient counts that either correlate with the edge of the SAA (typically $\geq 5\%$ pixel), or near the North/South terrestrial poles ($\sim 1-3\%$).
Given that the strength of the geomagnetic field in the SAA drops smoothly \cite{PavonCarrasco2016}, the elevated transient rates seen when SPHEREx approaches the SAA can be inferred as simply a limitation of using a location-based cutoff.
The moderately high rates at high North/South latitudes come from increased charged particle fluxes due to the shape of the Earth's magnetic field at the poles.

Secondly, detectors 1 to 3 
consistently have fewer transient counts than detectors 4 to 6.
Detectors 4 - 6 are 5$\micron$ cutoff H2RG arrays which have a lower Cadmium fraction in their photosensitive layers resulting in $\sim2\times$ lower effective energy band gap, compared to the 2.5$\micron$ cutoff H2RG arrays in detectors 1 - 3 \cite{Blank2012}.
Cosmic rays that deposit in more than one pixel in a single event (referred to as ``snowball'' in SPHEREx data) are observed to spread out to more pixels in detectors 4 - 6\cite{Fazar2026}, potentially reflect the lower well-depth of the 5$\micron$ arrays\cite{TeledyneBrochure}.

\begin{figure}[htbp]
	\centering
    \vspace{10pt}
	\includegraphics[width=0.8\textwidth]{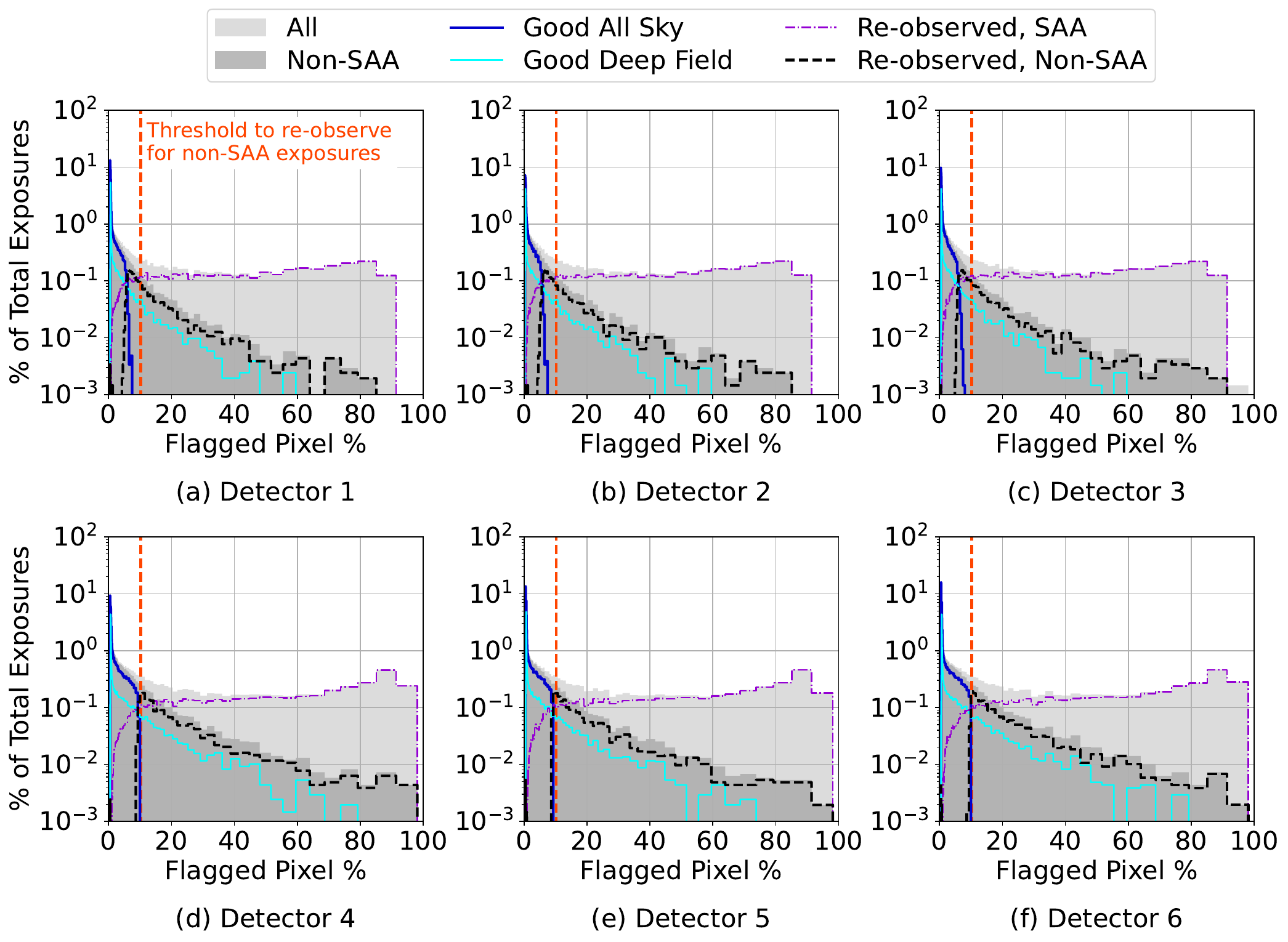}
    \vspace{5pt}
	\caption{Histograms of the transient rates from May 2025 through the end of April 2026 for six detectors. 
    The transient hit rate is reported as the percentage of flagged pixels relative to total pixel count in every array.
    To account for the dynamic range of the transient rates and that the majority of exposures have $< 2\%$ flagged pixels, the transient rate bins are uniform in logarithmic space.
    For each detector, the histogram of all exposures (shaded light gray bars) is decomposed into those labelled as SAA based on SPHEREx location (dot-dashed purple line) and non-SAA (shaded dark gray bars).
    The non-SAA data are further grouped into the deep field targets whose transient rates are not evaluated (solid cyan line), the all sky observations that are good for calibration (solid blue line) and those marked for re-observation (dashed black line).
    SAA observations make up the bulk of the exposures with hit rate above 10\%, with an almost uniform distribution.
    Comparing the non-SAA observations that are marked for re-observation (dashed black lines) in detectors 1 - 3 with detectors 4 - 6, the re-observation criterion appears to be mainly triggered by the latter.
	\label{fig:histogram}}
\end{figure}

The histograms of the transient hit rates in all 206,379 observations are plotted in Figure \ref{fig:histogram} for each detector, where we also decompose them into good (not flagged for re-observation) all sky, good deep field, and targets to be re-observed based on SAA flag and high transient rate.
The average transient rates of each group are listed in Table \ref{tab:sumstat} with $\pm1\sigma$ standard deviation.
The good deep/all sky observations here are those available to be calibrated, and later quality check may remove more exposures in these groups from the public archive (for example: exposures in crowded star fields can fail astrometry registration).
We find that SAA observations account for $\sim77\%$ of the high transient exposures.
Interestingly, the hit percentage distribution of the SAA is almost uniform.
In the case of the non-SAA re-observations, most re-observation flags are triggered by detectors 4 to 6.
While many exposures in detectors 1 - 3 have lower transient rates than the $10\%$ threshold, a re-observation is marked if any detector sees more than $10\%$ rate.
As seen in Figure \ref{fig:timeseries_zoomin}, the 5$\micron$ detectors consistently see more transient events than the 2.5$\micron$ detectors, and are the main driver behind the non-SAA re-observation flags.

\begin{table}
\begin{center}
\renewcommand{\arraystretch}{1.25}
\begin{tabular}{c|ccccc}
\hline
\hline
\multirow{2}{*}{Detector}  & \multicolumn{5}{c}{Mean Transient Hit as \% of Total Pixel per Exposure ($\pm1\sigma$)}\\
\cline{2-6}
                    &Good All Sky & Good Deep Field & Re-obs, SAA & Re-obs, non-SAA & All\\
\hline
1   & 0.90 (0.79) & 1.57 (2.98) & 29.70 (26.45) & 12.41 (9.69) & 3.00 (9.79)\\
2   & 0.77 (0.81) & 1.43 (3.04) & 29.64 (26.53) & 12.57 (9.99) & 2.88 (9.83)\\
3   & 0.91 (0.87) & 1.57 (3.10) & 29.84 (26.32) & 13.07 (10.49) & 3.03 (9.83)\\
4   & 1.19 (1.37) & 2.24 (4.56) & 39.99 (30.62) & 19.35 (13.04) & 4.09 (12.50)\\
5   & 1.13 (1.29) & 2.18 (4.35) & 39.07 (30.33) & 18.45 (12.48) & 3.97 (12.25)\\
6   & 1.17 (1.41) & 2.30 (4.67) & 40.74 (30.84) & 19.81 (12.96) & 4.14 (12.69)\\
\hline
\# of Exp & 151,623 & 38,359 & 12,697 & 3,699 & 206,378 \\
\% of Total & 73.47\% & 18.59\% & 6.15\% & 1.79\% & 100\%\\
\hline
\end{tabular}
\end{center}
\caption{Detector Mean Transient Hit per Exp \% ($\pm1\sigma$), divided into the survey fields and types of re-observation (rows 1 to 6).
The numbers of exposures (row 7) and the corresponding percentage values (row 8) are the same for six detectors.
\label{tab:sumstat}}
\end{table}

\begin{figure}[htb]
	\centering
    \vspace{10pt}
	\includegraphics[width=0.9\textwidth]{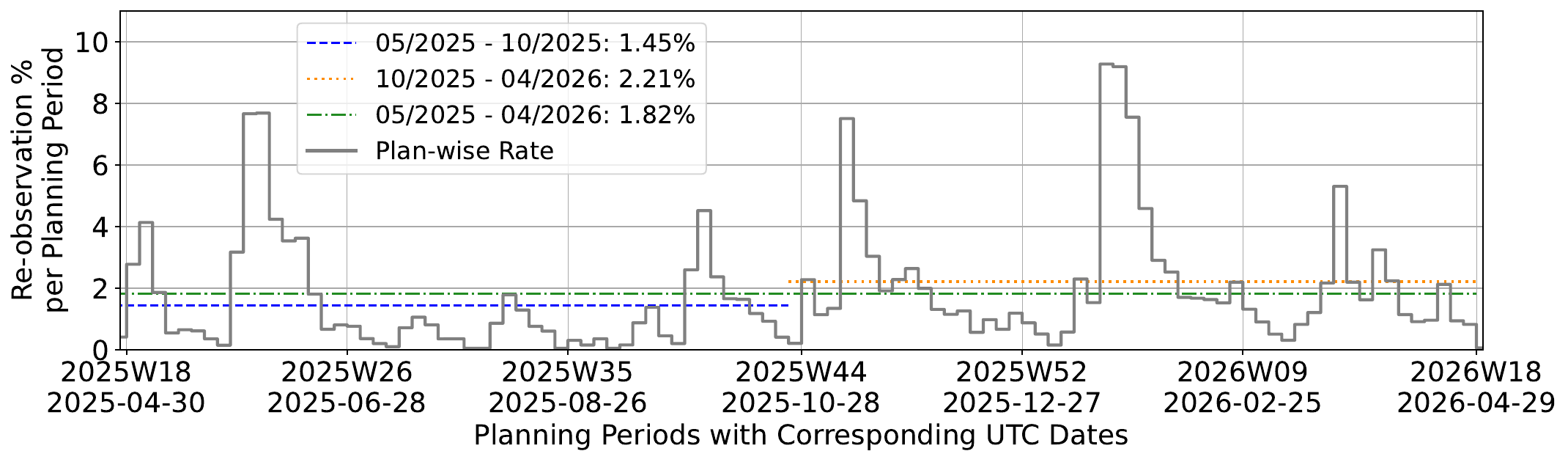}
    \vspace{3pt}
	\caption{The percentage of re-observed targets relative to the total number of observations against the biweekly 3.5-day planning periods, non-SAA observations only.
    The mean rate is 1.45\% in the first six months (May - October 2025, dashed blue line), increasing to 2.21\% in the last six months (November 2025 - April 2026, dotted orange line), for an effective first-year rate of 1.82\% (dot-dashed green line).
    The non-SAA re-observation rate appears to track the disturbances in the geomagnetic field, notably around early May/June/October/November of 2025, and January/February 2026.
	\label{fig:historical_reobs_rate}}
\end{figure}

To monitor the survey progress, we compute the non-SAA re-observation rate per planning period as the percentage of the re-observed sky targets relative to the total number of targets in the same period, and the result is plotted in Figure \ref{fig:historical_reobs_rate}.
The mean rate was 1.45\% in the first six months (May to October 2025), but went up significantly in the next six months to 2.21\% due to prolonged periods of geomagnetic disturbances, particularly in November 2025 and January to February 2026.
Nevertheless, the first year re-observation rate is $\sim1.82\%$, within the margin to complete two full-sky surveys in 12 months.

\section{THE SAA TRACED BY SPHEREx}\label{S:discussion}

\subsection{Two-Dimensional Map of the Transient Rates}
\begin{figure}[htb]
	\centering
    \vspace{10pt}
	\includegraphics[width=0.9\textwidth]{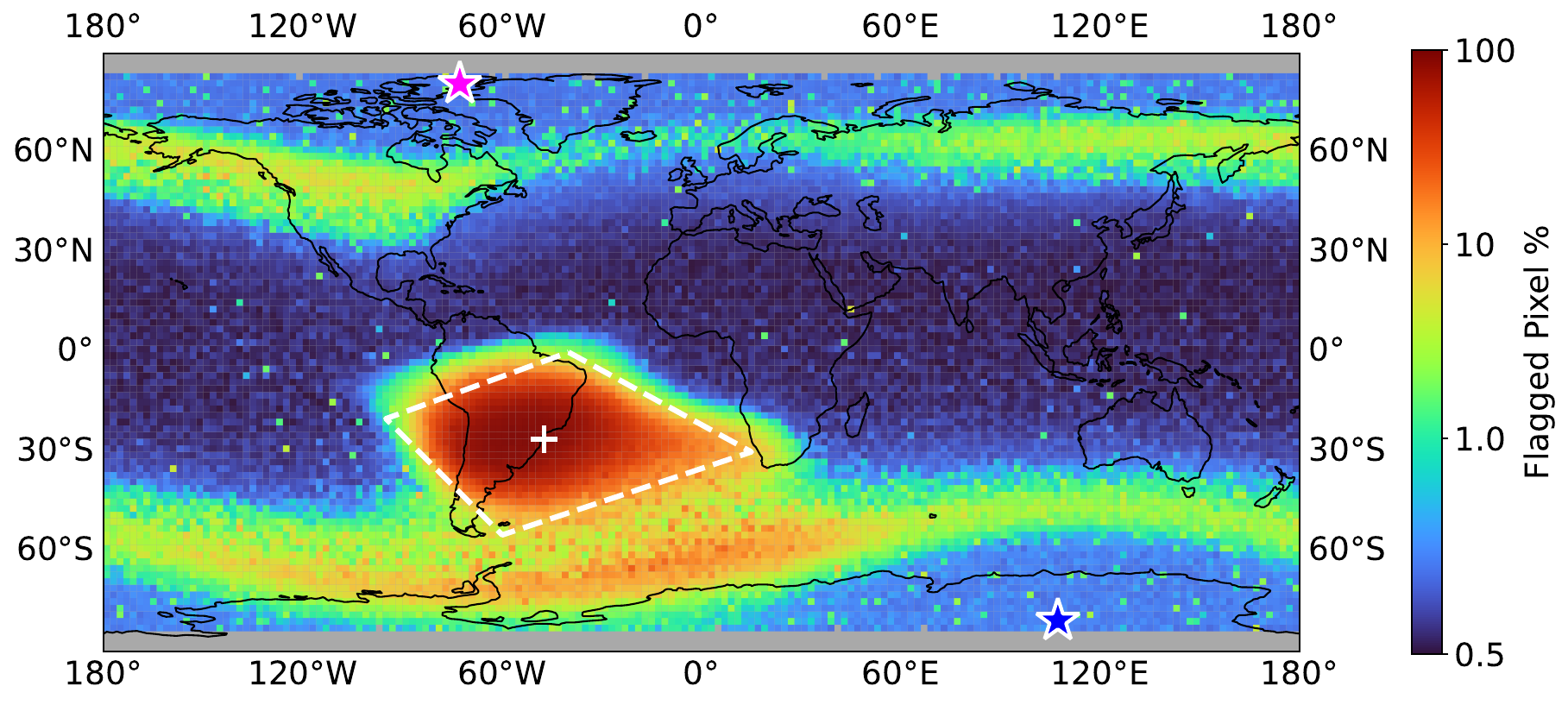}
    \vspace{5pt}
	\caption{The two-dimension distribution of the average transient hit rate from all six SPHEREx detectors.
    We exclude data from planning periods with $>2\%$ re-observation rates to capture only the shape of the SAA during quiet geomagnetic condition.
    The rate is binned in $2\degree$$\times$$2\degree$ longitude-latitude grid ($\sim$250$ \times$250$~\textrm{km}^2$ at SPHEREx altitude).
    The centroid of the transient hit rate is denoted by the white plus sign.
    The white dashed diamond shape represents the definition used by SPHEREx for SAA flagging.
    In addition to the SAA, SPHEREx transient trace two oval regions near the North and South geomagnetic poles, whose 2025 locations\cite{WMM2025} are marked by the magenta and blue stars respectively.
    SPHEREx map qualitatively agrees with previous results including the centroid location \cite{Bartocci2025, Ginisty2024, Ginisty2023}.
	\label{fig:heatmap_all}}
\end{figure}

Using the first year of data, we present in Figure \ref{fig:heatmap_all} the two-dimensional distribution of the average transient rate at every Earth position on a $2\degree$$\times$$2\degree$ grid ($\sim$250$ \times$250$~\textrm{km}^2$ at 670 km altitude).
To construct this map, we compute the average transient rate from all six detectors. 
While systematic offsets exist between the 2.5$\micron$ detectors (1 - 3) and 5$\micron$ detectors (4 - 6), the detector-to-detector variations are subdominant to the ground-tracked location variations as seen in Figure \ref{fig:timeseries_zoomin}.
To avoid the temporary enhancement from strong solar storms, we exclude data in planning periods with re-observation rate above 2\%.
The centroid of the SAA is computed from only the SAA observations as followed:

\begin{align}
\textrm{longitude}_\textrm{centroid} = \frac{\sum{\textrm{\# transient} \times \textrm{longitude}}}{\sum{\textrm{\# transient}}}\\
\textrm{latitude}_\textrm{centroid} = \frac{\sum{\textrm{\# transient} \times \textrm{latitude}}}{\sum{\textrm{\# transient}}}
\end{align}

For comparison, we overplot the boundary defined by SPHEREx operations team on Figure \ref{fig:heatmap_all}, which demonstrates that the location-based SAA flagging is sufficient to identify $\sim77$\% observations with high transient rates above 10\%.
However, the location-based flagging does not account for other regions with elevated transient rates, namely two oval regions centering on the North and South geomagnetic poles \cite{WMM2025} that are very clearly traced by the non-SAA transient rates.
Qualitatively, our map agrees with previous maps of the SAA using particle fluxes \cite{Bartocci2025, Ginisty2024, Ginisty2023}.
Additionally, we observe scattered clusters of high transient rates in the Southern hemisphere to the West of the SAA, distinctively away from the Southern geomagnetic pole.
These clusters appear to fall into a belt of low magnetic field strength previously mapped by ESA/Swarm data \cite{PavonCarrasco2016}, so they are likely similar indicators of regions of increased high-energy particle fluxes, except on a smaller scale than the SAA.

\subsection{Variations during Solar Storms}
\begin{figure}[htb]
	\centering
    \vspace{10pt}
	\includegraphics[width=0.9\textwidth]{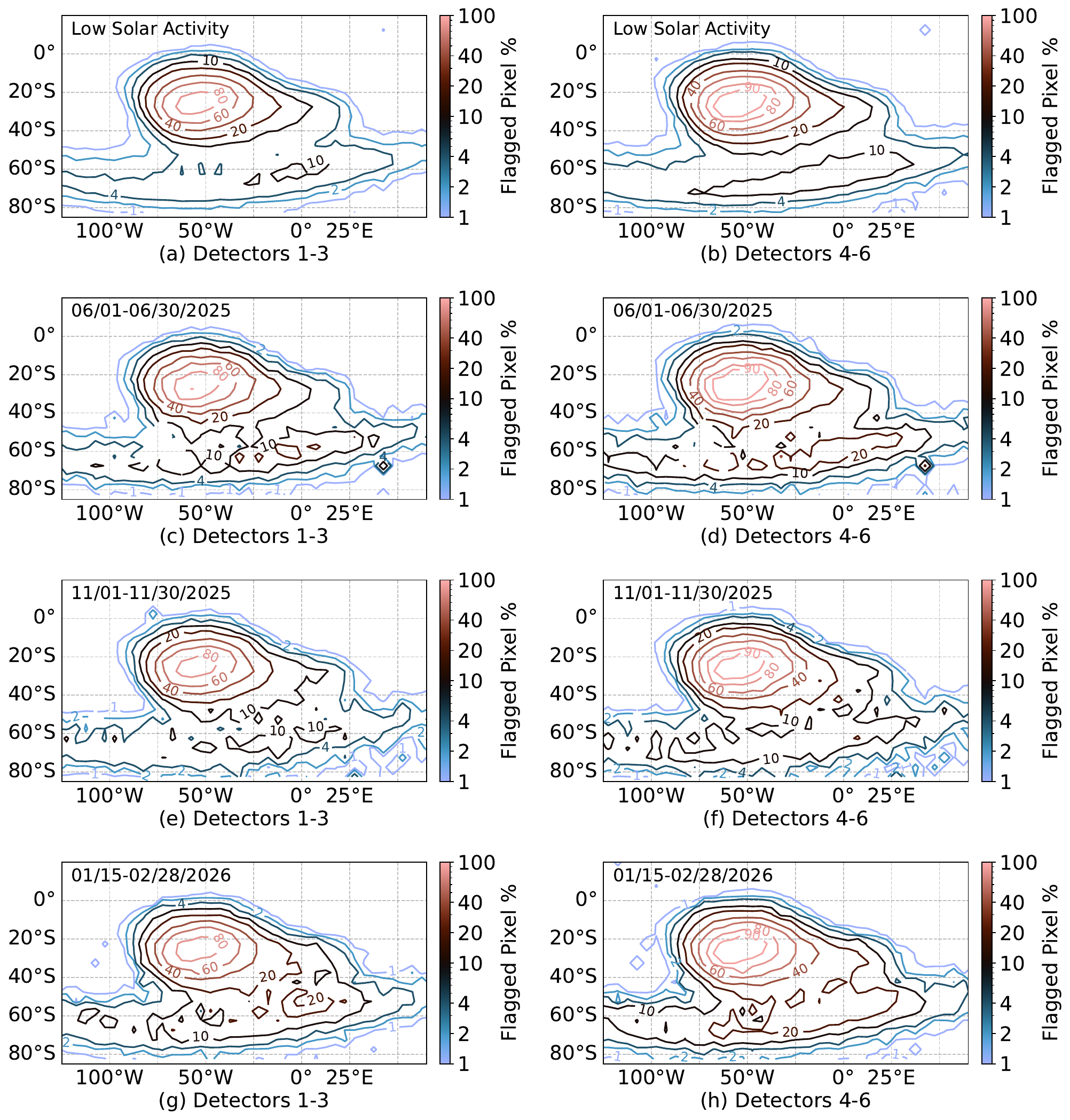}
    \vspace{5pt}
	\caption{Contour plots of the SAA traced by SPHEREx transient rates in various geomagnetic disturbance levels.
    To identify any changes below the detector-to-detector variations, we construct the contour maps separately for detectors 1 - 3 and detectors 4 - 6.
    (a)(c)(e)(g) are computed from the average transient rates of detectors 1 - 3, and (b)(d)(f)(h) from detectors 4 - 6.
    From top to bottom rows, the data correspond to low solar activity (similar to Figure \ref{fig:heatmap_all}), 1 - 30 June 2025, 1 - 30 November 2025, and 15 January to 28 February 2026 respectively.
    These three time periods are chosen due to the presence of strong solar storms leading to notably higher than usual re-observation rates as seen in Figure \ref{fig:historical_reobs_rate}.
	\label{fig:contours_all}}
\end{figure}

We examine the variations of the SAA shapes and boundaries in various geomagnetic disturbance levels in Figure \ref{fig:contours_all}.
To identify potential differences below the detector offset levels, we make two separate sets of maps for detectors 1 - 3 and detectors 4 - 6.
Furthermore, we increase the longitude-latitude grid size from $2\degree$$\times$$2\degree$ to $5\degree$$\times$$5\degree$ to account for fewer available ground-track positions in a one-month period.
We define low Solar activities similar to the previous section.
To contrast with this condition, we produce the maps of the SAA using only data in June 2025, November 2025, and mid-January to February 2026.
There are noticeable increases in the levels of transient hits during strong geomagnetic disturbances.
However, the 1\% and 2\% transient rate contours remain steady across different conditions.
Therefore, we conclude that the boundary does not vary in an appreciable manner during solar storms.

\section{SUMMARY}\label{S:summary}

SPHEREx is intended to be an astrophysics mission, but has been demonstrated to hold potential beyond astrophysics, for example through its ability to continuously detect and monitor airglow emission from helium and oxygen in the upper atmosphere \cite{Hui2026b}.
Similar to the airglow measurements, transient hit rate is a constant byproduct of the survey that can contribute to research in heliophysics and space weather.
In this paper, we reviewed the transient detection algorithm of SPHEREx, presented general trends of transient hit rates in one year of SPHEREx all-sky survey data, and demonstrated that transient rates can be used to map the SAA, an important feature in the geomagnetic field, as well as to probe temporal variations in different geomagnetic disturbance levels. 

Using the two-dimensional distribution of the transient rates across all detectors, we produce a map of the SAA during quiet geomagnetic conditions.
Aside from the SAA, the map reveals that SPHEREx detects elevated level of transients when passing over the geomagnetic poles as well as in multiple isolated regions in the Southern hemisphere, potentially probing the extended zone of the weakened geomagnetic field near the SAA.

Outside of the SAA and the SPHEREx deep fields (North/South ecliptic pole targets), on average an additional $\leq 2\%$ of observations see elevated transient rates that hinder the science calibration, so they are also removed from the data processing and the corresponding sky targets are marked for re-observation.
The re-observation rates roughly track Solar activity, where planning periods during major solar storms in 2025 and early 2026 saw significantly higher than 2\% of exposures exceeding the re-observation criteria.
Future work can correlate the temporal variations of the transient rates with fluctuation measurements of the geomagnetic disturbance tracers, for example the global Auroral Electrojet (AE) index, an indicator of geomagnetic substorms that cause particle precipitation into the upper atmosphere from the magnetotail \cite{Davis1966}.

\section*{ACKNOWLEDGMENT}
The authors acknowledge many extremely helpful discussions with Olga Verkhoglyadova and Panagiotis Vergados at the Jet Propulsion Laboratory, Katrina Bossert and Jessica Norrell at the Arizona State University, and members of the SPHEREx science and engineering teams.

We acknowledge support from the SPHEREx project under a contract from the NASA/Goddard Space Flight Center to the California Institute of Technology (80GSFC18C0011).
Part of the research described in this paper was carried out at the Jet Propulsion Laboratory, California Institute of Technology, under a contract with the National Aeronautics and Space Administration (80NM0018D0004).
The authors acknowledge the Texas Advanced Computing Center (TACC) at The University of Texas at Austin for providing computational resources that have contributed to the research results reported within this paper.
The High Performance Computing resources used in this investigation were provided by funding from the JPL Enterprise IT Services division.
This article and analysis made use of the Astropy, Numpy, Scipy, and Matplotlib Python packages. 

% References
\bibliography{ref} 
\bibliographystyle{spiebib} 

\end{document}